\documentstyle[12pt] {article}
\begin{document}
\thispagestyle{empty}
\begin{flushright}
UA/NPPS-9-98
\end{flushright}
\vspace{2.5cm}
\begin{center}
{\large {\bf Worldline Approach to Forward\\ and Fixed Angle
fermion-fermion Scattering\\ in Yang-Mills Theories at High Energies}\\}
\vspace{1.8cm}
A. I. Karanikas and C. N. Ktorides\\
\smallskip
{\it University of Athens, Department of Physics\\
Nuclear and Particle Physics Section\\
Panepistimiopolis\\
GR-15771 Athens, Greece\\}
\vspace{.2cm}
June, 1998
\end{center}
\vspace{1.2cm}
\begin{abstract}
Worldline techniques are employed to study the general behaviour of the
fermion-fermion collision amplitude at very high energies in a non-abelian
gauge field theory for the forward and fixed angle scattering
cases. A central objective of this work is to demonstrate the simplicity
by which the worldline methodology
isolates that sector of the full theory which carries the soft physics,
relevant to each process. Anomalous
dimensions pertaining to a given soft sector are identified and
subsequently used to facilitate the renormalization group running of
the respective four point functions. Gluon
reggeization is achieved for forward, while Sudakov behaviour is
established for fixed angle scattering.
\end{abstract}

\newpage 
\setcounter{page}{1}

{\large {\bf 1 Introduction}}

  The theoretical confrontation of collision experiments at high energies
calls for methodologies that have an essential dependence on the
kinematics
of the process. Fixing our ideas on investigations addressing themselves
to hadronic structure -even though references to QED, whenever relevant,
will also be made in this work- one of the key issues involved concerns
the interplay between the (invariant) center of mass energy $\sqrt{s}$, on
the one hand, and the (invariant) momentum transfer $\sqrt{-t}$, on the
other. Asymptotic regimes corresponding to the cases: (a)$|t|$ large
-${t\over 2pq}$, $p$ proton and $q$ photon four-momentum,
fixed (Bjorken limit) and (b) $s$ large -$t$ fixed (Regge limit) have come
under extensive theoretical scrutiny, especially in connection with deep
inelastic scattering (DIS) processes.

Given that momentum transfer defines the resolution by which the short
distance structure of a hadron (nucleon) is being probed, the Bjorken
limit has naturally taken historical precedence in both the experimental
and the theoretical front. More recently, the Regge kinematical regime has
been receiving wide attention in view of the ongoing experiments at HERA.
When the emphasis on asymptotics shifts to $s$, it is the hadronic profile
imprinted onto the plane transverse to the direction of the collision that
forms the basis of theoretical interest. In a perturbative context, which
one readily adopts by taking $-t \gg \Lambda^2_{QCD}$, the emerging
picture is that of a high density distribution of partons accross the
surface perpendicular to the motion, each carrying a small fraction $x(\ll
1)$ of the
hadron momentum. In such a context the dynamics pertaining to the process
is studied in the two dimensional transverse plane where, once again, 
momentum transfer facilitates the probing of the hadron. The 
situation just described fits exactly into an eikonal mode of description
wherein the
no impulse approximation forces the momentum of the exchanged quanta to
have (appreciably) non-vanishing components only in the transverse
direction.

The most systematic quantitative considerations, referring to the 
qualitative account given above in connection with the $s-t$ interplay, 
have been made, within the framework of perturbation
theory, by Cheng and Wu [1] for both abelian and non-abelian gauge systems
with spin-1/2 matter fields. Special emphasis, in the work of these
authors, was placed
on the issue of unitarity which calls for separate attention to the $s$
and the $t$ channel, respectively. The so called Cheng-Wu towers, in QED, 
which stretch along the $t$ (vertical) direction present, upon cutting, a
fragmented profile of the electron (positron), see Fig. 1a. In non-abelian
theories, on
the other hand, Cheng and Wu discover reggeization of the gluon exchange
 among the colliding particles. The corresponding
dominant towers that unfold along the $t$-channel are formed by the
exchange of gluons between reggeons, see Fig. 1b.

In addition to the Bjorken and Regge limits a third situation which
presents interest, all of its own, is defined by the specification $s,\,
|t|\rightarrow\infty$ at fixed ratio $s/t$. We shall henceforth refer to
it
as fixed angle kinematical regime, implying that the angle is fairly wide
and that the collision energy is very large. With respect to DIS we expect
this case to be relevant in semi-inclusive processes, when the observed
particle in the final state emerges at an angle with respect to the
(virtual) photon-nucleon direction.

Having said the above let us define the bounds of the present work by
specifying that what we intend to pursue is the problem of the non-abelian
scattering among fundamental spin-1/2 fields, cf. isolated quarks, in the
Regge and
fixed angle kinematical regimes. Our main objective is to show how the
worldline casting of field systems, which we have been systematically
pursuing in recent years [2-6], leads to efficient and straight forward
methods of calculation, considerably simpler than corresponding procedures
developed within the Feynmann diagrammatic description of QCD. We place
particular significance
on the issue of factorization between soft and hard physics entering a
given process of interest which, next to confinement, is the most
important aspect of QCD applications -certainly the one that lends itself
to present day capabilities in coping with the theory.

It becomes obvious that, in this work, we shall neither venture into the
domain of (exclusive) hadron-hadron scattering, where
phenomenologically-based factorization issues [7-10] involving hadronic
wave functions [11,12] become essential, nor into DIS matters
where structure functions and evolution equations [13,14] assume primary
role.
Clearly, such undertakings pose ultimate goals of pursuit and 
set directions for future work. Our main concern, presently, is the 
identification of {\it global} behaviors -as opposed to 
investigations of structure-
underlying each of the amplitudes we are interested in, i.e. what one  
naturally associates with the soft physics. 

The basic advantage of the worldline casting of a theory, such as QCD, is
the spacetime setting that underlines the scheme as a whole [15-17,2,3].
This facilitates considerations which, being much
closer to intuitive, geometrical pictures, as opposed to what one can
associate with Feynman diagrams, lead to efficient computational
procedures [16-18,3-6]. Considerations of similar nature have already
been evidenced in the work of several authors in connection with the study
of Wilson loops [19-23] which, after all, correspond to (Euclidean) 
``worldline" contours of
infinitely heavy matter fields. The difference, at a foundational level,
is that in our case Wilson loops/lines enter as natural ingredients of the
field theoretical description of the system {\it per se} -see, e.g., [4]-
and not as part of an operator-based formalism -see,e.g., [24,25]. A
practical
consequence of this occurence is our ability to deal with Wilson line
operators of {\it finite} length. 
As shown in [6], manipulations with {\it open Wilson lines} (of finite
extent) in worldline formalism when applied to forward QED processes are
naturally associated with
off mass shell eikonals. We intend to capitalize on this fact in the
present paper using it to achieve infrared regularized expressions by
going off mass shell.  

Renormalization issues will also play a central role in our subsequent 
analysis. As already
alluded to above, our main preoccupation is to isolate a sector of the
full Yang-Mills field system wherein the active degrees of
freedom are the soft ones. Even within such a subsector, the disparity
between its upper momentum cutoff and the infrared one\footnote{It is
important to distiguish between what one terms `soft' and what
`infrared': The first characterization pertains to {\it observable}
degrees of freedom while the second refers to unobservable ones which
presumably reside in the `vacuum' state.} gives rise to
anomalous dimensions, having appropriate interpretations.  Our first
illustration of the situation in hand
will be presented in connection with the forward amplitude, in the
$s\rightarrow \infty$, $t$ fixed kinematical region. An immediate
application of the resulting structure will be the establishement, in the
LLA, of the reggeization of non-abelian gauge fields exchanged among the
colliding fermions. This will be readily accomplished via a
renormalization group running within the soft subtheory, which employs
the abovementioned anomalous dimensions [26].

 A more demanding task is posed by the fixed angle scattering process. The  
bulk of our efforts in this case will fall upon factorization issues which
become more compelling, in comparison to the previous (Regge) one, due to
the presence of a hard momentum scale. The idea is to first identify the
relevant anomalous dimensions in the, factorized, soft sector and then
proceed to derive the expression for the whole amplitude, in the LLA,
exploiting its invariance with respect to the scale which separates soft
from hard components. As it
turns out, the end result of this analysis is the emergence of a
Sudakov [27,28] suppression factor which dampens the four-point process.

The particulars of our worldline casting of field systems have been
extensively discussed in several papers [2-6], so the interested reader
should refer to these sources, especially [3]. Suffice it to say that such
a casting amounts to a reformulation $\int {\cal
D}\overline{\psi}(x)\,{\cal D}\psi (x)\,
e^{S[\overline{\psi}(x),\psi (x), A_\mu (x)]}...\longrightarrow \int {\cal
D}x(\tau)\,{\cal D}p(\tau)\,e^{S[x(\tau),p(\tau),A_\mu(x(\tau))]}...$, 
taking us from a functional to a path integral description of the system.
Note that the above transcription pertains to the fermionic sector of the
gauge field theory which registers via quadratic terms in the action.
Accordingly, nothing is lost as one carries out the Gaussian-Grassmannian
integrations over the corresponding fields. Functional integration with
respect to the gauge fields remains to be carried out and it is within
this context that the dynamics operating in the system reveals itself.

 The organization of our paper is straight forward: In Section 2 we deal
with Regge limit behaviour, while in Section 3 we study the fixed angle
kinematical 
regime for (isolated) fermion-fermion scattering in a non-abelian gauge
field system. (The extraction of two expressions entering the soft part of
the fixed angle amplitude is traced in an Appendix.) For the convenience
of the reader each section is divided
into subsections, addressing respective issues in a self-contained manner. 
A final, brief, section is devoted to conclusions and outlook.

\vspace*{.4in}
{\large {\bf 2 Non-abelian Scattering Amplitude at Regge Asymptotics}}

In this section we shall consider the scattering of two spin-1/2 particles
belonging to the fundamental representation of a given non-abelian gauge,
$SU(N)$, group in the limit $s\rightarrow
\infty,\,|t|(\gg\Lambda^2_{QCD})$ fixed. As already stated, we shall
employ
a worldline mode of description for the process in the context of which
the eikonal approximation acquires a sharp, geometrically-based,
interpretation. Our efforts will address the amplitude directly. Special
emphasis will be placed on our ability to control IR divergencies by going
off-shell whereupon one deals with open Wilson line operators\footnote
{By contrast, the on mass-shell alternative, which calls for an explicit
gluon mass as IR regulator, corresponds to the employment of Wilson {\it
loop} operators.}. A renormalization group running which leads to
gluon reggeization will be carried out, while some specific remarks
pertinent to QED will be made in the end.

\vspace*{.2in}
{\bf 2.1 Four-point function in worldline formalism}

Working in Euclidean space we introduce the four-point function
\begin{eqnarray} 
\frac{\delta
^4}{\delta\overline{\eta}^f_i(x_1)\,\delta\overline{\eta}^f_j(x_2)\,\delta 
\eta^f_{i'}(y_1)\,\delta\eta^f_{j'}(y_2)}&&lnZ(\overline{\eta},\eta)
|_{\overline{\eta}=\eta=0} \nonumber\\     &=& {\cal
M}(x_1,x_2;y_1,y_2)^{ii'}_{jj'}\,-\,{\cal M}(x_1,x_2;y_1,y_2)^{ij'}_{ji'},
\end{eqnarray}
where $Z(\overline{\eta},\eta)$ is the partition function for the
non-abelian gauge field theory with spin-1/2 matter fields whose sources
are 
$\overline{\eta},\,\eta$, while $f$ is a flavor and  $i,j...$ are group  
representation indices. The two, connected, four-point amplitudes on the
right hand side are related via a particle exchange in the final state.
The second of the two terms, however, gives little contribution to the
forward direction scattering process, which presently interests us, so its
presence will be ignored for the rest of this section.

The expression for the amplitude in worldline formalism is

\begin{eqnarray}
{\cal M}^{ii'}_{jj'}\, =\,\displaystyle{\sum_{C^I_{x_1,x_2}}}\,
\displaystyle{\sum_{C^{II}_{x_2,y_2}}}\,I[\dot{x}^I]\,I[\dot{x}^{II}] & &
<Pexp[ig\int_0^{T_1}\,d\tau\,\dot{x}^I(\tau)\cdot A(x^I(\tau))]_{ii'}
\nonumber\\ & & \times Pexp[ig\int _0^{T_2}\,d\tau
\dot{x}^{II}(\tau)\cdot A(x^{II}(\tau))]_{jj'}>_A^{conn},
\end{eqnarray}
where
\begin{eqnarray}
\displaystyle{\sum_{C^I_{x,y}}}I[\dot{x}]\,\equiv\,& &\int_0^\infty dT
\displaystyle{\int_{\stackrel{x(0)=x}{x(T)=y}}}\,{\cal D}x(\tau) \int
{\cal
D}p(\tau)\,Pexp[-\int^T_0\,d\tau(ip(\tau)\cdot\gamma\,+\, m_f)]\nonumber\\
& &\times exp[i\int_0^T\,d\tau\,p(\tau)\cdot \dot{x}(\tau)].
\end{eqnarray}

Notice that the spin-1/2 sector of the theory enters our expressions
through one dimensional geometrical contours (Euclidean ``worldlines"),
while all the dynamics is contained in the expectation values
$<\cdot\cdot>_A^{conn}=<\cdot\cdot>_A-<\cdot>_A<\cdot>_A$ of
Wilson line operators\footnote{We denote by $A$
the gauge field expanded in the Lie algebra. Also, our compressed notation
implies that an expectation value in the gauge field sector includes gauge
fixing terms, ghost integration and, in the generic case, contribution
from the Dirac determinant.}.

The spacetime setting of the worldline approach affords us to designate
points $z_1$ and $z_2$, on the respective contours $x^I(\tau)$ and
$x^{II}(\tau)$, of closest approach. Setting $x^I(s_1)\equiv z_1,\,
x^{II}(s_2)\equiv z_2$ and using the identity
$\int_0^T\,\frac{ds}{T}\,\int\,d^4z\,\delta[x(s)-z]=1$ we write

\begin{eqnarray}
{\cal M}^{ii'}_{jj'}\,=\,\displaystyle{\sum_{C^I_{x_1,y_1}}}
\displaystyle{\sum_{C^{II}_{x_2,y_2}}}\,I[\dot{x}^I]\,I[\dot{x}^{II}] & &
\int_0^{T_1} \,\frac{ds_1}{T_1}
\int_0^{T_2}\,\frac{ds_2}{T_2}\int\,d^4z_1\,\delta[x^I(s_1)-z_1]\nonumber\\ 
& & \times\int\,d^4z_2
\,\delta[x^{II}(s_2)-z_2]\, E(C^I,
C^{II};z_1,z_2)^{ii'}_{jj'},
\end{eqnarray}
where\footnote{Notice the notational shortcut: $\int_s^T
d\tau\dot{x}(\tau)\rightarrow \int_s^T\dot{x}$ inside the arguments of the
$A$'s.}
\begin{eqnarray}
& & E(C^I,C^{II};z_1,z_2)^{ii'}_{jj'}=\nonumber\\ & &
<Pexp[ig\int_0^{s_1}d\tau
\dot{x}^I(\tau)\cdot A(z_1-\int_\tau^{s_1}\dot{x}^I)
+ig\int_{s_1}^{T_1}d\tau \dot{x}^I\cdot A(z_1+\int_{s_1}^{T_1}
\dot{x}^I)]_{ii'} \nonumber\\ & &\quad \times Pexp[ig\int_0^{s_2} d\tau 
\dot{x}^{II}\cdot A(z_2-\int_\tau^{s_2}\dot{x}^{II}) +
ig\int_{s_2}^{T_2}d\tau\dot{x}^{II}\cdot A(z_2+
\int_{s_2}^{T_2}\dot{x}^{II})]_{jj'}>_A^{conn}.
\end{eqnarray}

The above battery of expressions contains the basic formalism of the
worldline approach to non-abelian gauge systems with spin-1/2 matter
fields pertaining to the four-point process. Our adjustments, from hereon,
will refer to the particular situations, i.e. Regge and fixed angle
kinematics we intend to study in this and the next section, respectively. 

\vspace*{.2in}
{\bf 2.2 Eikonal approximation in the worldline formalism}

Our main applications of the worldline scheme have addressed themselves to
the issue of isolating a subsector of a given microscopic theory which can
be characterized as ``soft". By the latter term we mean a restriction to
the study of the physics which is active in the field theory 
below a given scale $\tilde{\Lambda}$. Clearly, the full
microscopic theory contains degrees of freedom associated with higher
frequencies. Our strategy, however, is to
incorporate them into the definition of the physical quantities
entering the theory at scale $\tilde{\Lambda}$ (bare values). For the
purposes of this section the aformentioned scale could correspond either 
to physical mass of the matter field quanta \footnote{We resrtict
ourselves to a single physical mass parameter $m$. Adjustments pertaining
to different masses could be made at the
expense of burdening our analysis with extra formalism that would detract
from our main objectives.} (on-mass-shell situation) or to an off-shell
mass value.

 The beauty of the
worldline casting of the field system is that it effects the isolation of
the ``soft" subsector in a most efficient and straight forward manner: One
instructs the path integral to take into consideration only those paths
that are straight lines almost everywhere (allowing, therefore, for the
presence of cusps) and sets the Dirac
determinant to unity. In physical terms the above specifications imply
that matter fields have been dressed to the point that the live, in the
considered subsector of the full theory, gauge field
exchanges can neither derail them from their propagation paths nor create
virtual pairs from the vacuum. Any derailment occurs on a sudden impulse
basis and corresponds to the presence of cusps on the propagation contour.

Now, the ``soft" subsector has its own UV and IR domains. The latter
presumably coincides with that of the full theory, while the former
provides anomalous dimensions which induce renormalization group running
of physical quantities. Renormalization factors in the subtheory are
exclusively associated with the (almost everywhere)  straight line
configurations  and depend solely on the number of cusps that a given
contour, relevant to the situation being studied, has. In particular, a
(single) straight line of propagation from an
initial to a final space-time point signifies negligible momentum transfer
to the matter particle. In such a case a wave function renormalization
factor is all that is found to be associated with the contour. A cusp, on
the other hand, implies a transfer of momentum to the propagating matter
particle which occurs on a sudden impulse basis. One is then faced with
the task to renormalize the vertex that forms at the derailment point.

The above general comments, offered as a way to
provide a first feeling concerning renormalization aspects of the
subtheory which has been isolated by our aformentioned stipulations, will
receive quantitative treatment throughout our analysis. For the moment,
let us consider the case of a four-point process involving two straight
line paths $L^I$ and $L^{II}$ characterized by four-velocities $u_1
(=\dot{x}^I)$ and $u_2(=\dot{x}^{II})$, respectively. Let the closest
point
of approach be $|z|(=|z_1-z_2|)<\sigma$, where $\sigma$ represents the
(finite) length of the worldline paths and is of the order of the inverse
off-mass-shell momentum scale. The exchanged gauge field
quanta are assumed to be too soft to derail the two matter particles,
propagating along $L^I$ and $L^{II}$, in any appreciable way.
This situation is precisely what one encounters when studying high energy
forward scattering in the eikonal approximation and is depicted in Fig.2a.
We have already
considered the applicability of this scheme to QED as well as a linear
version of quantum gravity [6]. As already stated our concern, in this
paper, pertains to non-abelian gauge systems (QCD).

In the eikonal frame of description, as specified above, Eq (5) reads (we
set $|z|=z$ for economy)  
\begin{eqnarray}
E(L^I,L^{II};z)^{ii'}_{jj'} &=&\,<Pexp[ig\int^{+\sigma}_{-\sigma}d\tau\,
u_1\cdot A(\tau u_1)]_{ii'}\nonumber\\ & &\times
Pexp[ig\int^{+\sigma}_{-\sigma}d\tau\, u_2\cdot A(z+\tau
u_2)]_{jj'}>_A^{conn}.
\end{eqnarray}
The above expression furnishes, in the worldline scheme, the dynamical
factor which enters the amplitude, cf. Eq (4), and will serve as the
central piece of attention in our subsequent analysis.

\vspace*{.2in}
{\bf 2.3 Off mass shell IR regularization and renormalization issues}

It is clear that the quantity given by (6) is UV safe\footnote{We are
referring strictly to exchanges {\it between} the two lines (connected
four-point function).} as
long
as ${1\over z}<\infty$ whereas it is protected from IR singularities if
$\sigma<\infty$. The latter specification corresponds to an off mass shell
description of the matter particle as the finite length of its propagation
contour cuts off all gauge field modes with momentum less than $1/\sigma$
which participate in its full, on mass shell description.

We shall proceed to investigate the behaviour of the dynamical factor
$E^{ii'}_{jj'}$ by controlling the IR divergencies through line contours
of finite length and, following Korchemsky [26], consider UV implications 
as $z\rightarrow 0$. The relevant singularity is designated as `cross
singularity' since it arises at a point where the two worldlines cross
each other. Employing dimensional regularization for controlling this
cross UV divergence, a mass scale $\mu$ is introduced which, for {\it
finite} $z$ now, implicates a renormalization group running up to the
scale $1/z$. Accordingly, our primary task is to determine the anomalous
dimensions which will enter the renormalization group equation. 

Given the presence of path
ordered exponentials, entering on account of the non-abelian setting, our
only option is to proceed perturbatively. We write

\begin{equation}
E^{ii'}_{jj'}\,=\,
(ig)^2t^\alpha_{ii'}t^\alpha_{jj'}\int^{+\sigma}_{-\sigma}d\tau
\int^{+\sigma}_{-\sigma}d\tau '\, u_1\cdot u_2 D(|u_1\tau -u_2\tau '|) \,
+\, {\cal O}(g^4),
\end{equation}
where, employing the Feynman gauge, one has
\begin{equation}  
D(|x|)\, =\, \mu^{4-D}\int\frac{d^Dk}{(2\pi)^D}\,e^{-ik\cdot
x}\frac{1}{k^2}\, =\, \frac{\mu^{4-D}}{4\pi^{D/2}}\,\Gamma
\left(\frac{D}{2}-1\right)\frac{1}{|x|^{D-2}} \, .
\end{equation}
We easily determine that

\begin{equation}
\int^{+\sigma}_{-\sigma}d\tau\int^{+\sigma}_{-\sigma}d\tau '\,u_1\cdot
u_2\,D(|u_1\tau -u_2\tau '|)\, =\,
\frac{1}{4\pi^2}\,\left(\frac{\mu^2}{\tilde{\lambda}^2}\pi\right)^\epsilon 
\,\frac{1}{2\epsilon}\, f_{4-2\epsilon}(w),
\end{equation}
where $w\equiv u_1\cdot u_2$, $\tilde{\lambda} \equiv 1/\sigma$,
$\epsilon=4-D(>0)$ and
\begin{eqnarray}
f_{4-2\epsilon} &=& 4w\left[\sqrt{\pi}\Gamma
(\frac{1}{2}-\epsilon)(1-w)^{\epsilon
-\frac{1}{2}}-\frac{1}{2}(1+w)^\epsilon\right.
F\left(1,1-\epsilon;\frac{3}{2}-\epsilon;\frac{1-w}{2}\right)\nonumber\\
 & & -\frac{1}{2}(1-w)^\epsilon
\left.F\left(1,1-\epsilon;\frac{3}{2}-\epsilon;
\frac{1+w}{2}\right)\right].
\end{eqnarray}
Setting $w=cos\theta$ we obtain, in the limit $\epsilon\rightarrow 0$,
\begin{equation}
f_4(\theta)\,=\,2\pi cot\theta
\end{equation}
and in Minkowski space $(\theta\rightarrow -i\gamma)$
\begin{equation}
f_4(\gamma)\,=\, 2\pi i\,coth\gamma.
\end{equation}

Subtracting the pole term in (7), using the $\overline{\mbox{MS}}$
scheme, we write
\begin{equation}
(E_1)^{ii'}_{jj'}\, =\, (ig)^2t^\alpha_{ii'}t^\alpha_{jj'} 
\frac{1}{4\pi^2}[ln\left(\frac{\mu ^2}{\tilde{\lambda}^2}\right)\,
+h(\gamma)] i\pi coth\gamma,
\end{equation}
where
\begin{equation}
h(\gamma)\,\equiv \, \frac{1}{i\pi coth\gamma}lim_{\epsilon\rightarrow 0}
\frac{1}{2\epsilon}[f_{4-2\epsilon}(\gamma)\,-\,f_4(\gamma)].
\end{equation}.

In the limit $\gamma\sim s/m^2\rightarrow\infty$, where $m$ stands for the
single, according to our agreement, fermion mass scale one obtains
\begin{equation}
h(\gamma)\,=\,ln(s/m^2)
\end{equation}
It is important to realize that $h(\gamma)$ has entered our analysis on
account of the off mass shell procedure we are currently pursuing.

Taking into consideration the fact that the $t$-matrices are in the
fundamental representation we finally get
\begin{equation}
(E_1)^{ii'}_{jj'}\,=\,c_{11}\delta_{ii'}\delta_{jj'}\,+\,
c_{12}\delta_{ij'}\delta_{ji'},
\end{equation}
where
\begin{equation}
c_{11}\,=\,\frac{\alpha_s}{2\pi}ln(M^2/\tilde{\lambda}^2)\frac{1}{N}\,i\pi
coth\gamma\,+\,{\cal O}(\alpha^2_s)
\end{equation}
and
\begin{equation}
c_{12}\,=\,-\frac{\alpha_s}{2\pi}ln(M^2/\tilde{\lambda}^2)\, i\pi
coth\gamma
\,+\,{\cal O}(\alpha^2_s)
\end{equation}
with $M^2\equiv \mu^2e^{h(\gamma)}$.    

\vspace*{.2in}
{\bf 2.4 Reggeization of exchanged gluons}

The UV structure that has emerged from our considerations in the previous
subsection has produced, to ${\cal O}(\alpha_s^2)$, anomalous dimensions
of the form
\begin{equation}
\Gamma_{cross}\,=\,\frac{\alpha_s}{\pi}(-\frac{i\pi}{N}coth\gamma \, , \,
i\pi coth\gamma).
\end{equation}

As Brandt et al. [20] have already pointed out, {\it albeit} within the
context of a Wilson loop analysis (see next subsection), under
renormalization group running quantitity $(E_1)^{ii'}_{jj'}$ mixes with
\begin{eqnarray}
(E_2)^{ii'}_{jj'}&=&<Pexp\left[ig\int_{-\infty}^0 d\tau\,u_1\cdot A(\tau
u_1) \,+\, ig\int_0^{+\infty}d\tau\,u_2 \cdot A(\tau
u_2)\right]_{ij'} \nonumber\\ & &
\times Pexp\left[ig\int_{-\infty}^0d\tau u_2\cdot A(\tau u_2) \,+\, ig
\int_0^{+\infty}d\tau u_1\cdot A(\tau u_1)\right]_{ji'}>_A\,-\delta_{ij'}
\delta_{ji'}.
\end{eqnarray}
Notice that the worldline configuration entering the above expression
consists of two independent cusped lines (see Fig.2b). The two cusps face
each other
and constitute sources of bremsstrahlung emission. We shall have more
comments to make on this matter in the next subsection. For now let us
turn our attention to renormalization issues associated with this quantity
which are, this time, indigenous on account of the momentum transfer that
accompanies each cusp and which can be unboundedly large.

For computational convenience we choose to simulate on mass shell
regularization methodology
for confronting the IR divergences, introducing for this purpose a
(small) mass $\lambda\sim\tilde{\lambda}\sim{1\over\sigma}$ for the gauge
field quanta. This simplifies the computation, which now involves four
Wilson lines, as we are able to 
assign unit magnitude to each four-velocity. Consistency with our off
mass shell IR regularization {\it can}, on the other hand, be achieved by
employing the energy dependent renormalization scale $M^2\equiv\mu^2e^{h
(\gamma)}$ identified in the previous subsection.    

We readily determine, after making the necessary readjustment in order to
attain a result compatible with the $\overline{\mbox{MS}}$ subtraction
scheme,
\begin {eqnarray}
& & (E_2)^{ii'}_{jj'}=\nonumber\\& & 
2(ig)^2c_F\delta_{ij'}\delta_{ji'}\left[\int_0^\infty   
d\tau\int_0^\infty d\tau 'D(|u\tau -u\tau '|;M) +\int_0^\infty
d\tau \int_0^\infty d\tau ' D(|u_1\tau +u_2\tau '|;M) u_1\cdot
u_2\right]\nonumber\\
& &+ 2(ig)^2t^\alpha_{ij'}t^\alpha_{ji'}\left[\int_0^\infty
d\tau\int_0^\infty d\tau 'D(|u\tau +u\tau '|;M) 
+\int_0^\infty d\tau\int_0^\infty d\tau ' D(|u_1\tau-u_2\tau
'|;M)u_1\cdot u_2\right]\nonumber\\& & + {\cal O}(g^4),
\end{eqnarray}
where $u$ stands, where it appears, generically for $u_1$ and $u_2$ and
\begin{equation}
D(|x|;M)\, =\, M^{4-D}\int\frac{d^Dk}{(2\pi)^D}\, e^{-ik\cdot
x}\,\frac{1}{k^2+\lambda^2}\, .
\end{equation}
Once more we emphasize that the seemingly on mass shell regularization
implied by the above formula is adjusted to the off mass shell strategy,
that we have been adhering to, via the use of the energy-dependent
mass $M$ as our renormalization point. 

In the limit of asymptotically high energies, we obtain, in Minkowski
space, 
\begin{equation}
(E_2)^{ii'}_{jj'}\, =\,
\delta_{ii'}\delta_{jj'}c_{21}\,+\,\delta_{ij'}\delta_{ji'}c_{22},
\end{equation}
where
\begin{equation}
c_{21}\,=\, -\frac{\alpha_s}{2\pi}ln(M^2/\lambda^2)[\gamma coth\gamma +
i\pi coth\gamma]\, +\,{\cal O}(\alpha_s^2)
\end{equation}
and
\begin{equation}
c_{22}\,=\, -\frac{\alpha_s}{2\pi}ln(M^2/\lambda^2) [N(coth\gamma-1)
-\frac{i\pi}{N}coth\gamma]\,+\,{\cal O}(\alpha_s^2).
\end{equation}

From the above relations we read the anomalous dimensions associated with
the `pair cusp' configuration as follows
\begin{equation}
\Gamma_{pair\,cusp}=\frac{\alpha_s}{\pi}(-\gamma coth\gamma+1+i\pi
coth\gamma,\, N(coth\gamma-1)-\frac{i\pi}{N}coth\gamma).
\end{equation}

Combining the above result with that of Eq. (19) we obtain the following 
$2\times 2$ anomalous dimension matrix 
\begin{equation}
(\Gamma_{ab})=\frac{\alpha_s}{\pi}\left(\begin{array}{ll}
       -\frac{i\pi}{N}coth\gamma \, &\, i\pi coth\gamma\\
       -\gamma coth\gamma+1+i\pi coth\gamma\,&\, N(coth\gamma-1)-
\frac{i\pi}{N}coth\gamma \end{array}\right)
\end{equation}
which governs the
running of the quantities $E_1$ and $E_2$ under the renormalization group
equation. The generic form of the latter is
\begin{equation}
(M\frac{\partial}{\partial M}+\beta(g)\frac{\partial}{\partial g})
E_a\,=\,\Gamma_{ab}(\gamma,g)E_b.
\end{equation}

At this point we have established full contact with Korchemsky's
operator-based analysis. Following Ref. [26], we introduce amplitudes
$T_{\stackrel{+}{-}}$ which enter expressions for the singlet
and octet components of the invariant amplitude whose LLA form is given,
according to the renormalization group equation (28), by
\begin{equation}
T_{\stackrel{\textstyle +}{-}}=\frac{t}{\Gamma_{\stackrel{\textstyle
+}{-}}}\,\int d^2z\,e^{-i\vec{z}\cdot\vec{q}}\, exp\left
[-\Gamma_{\stackrel{\textstyle +}{-}}\int_\lambda^{\frac{1}{\bar{z}}}
\frac{d\tau}{\tau}\,\frac{\alpha_s(\tau)}{\pi}\right ],
\end{equation}
where $\Gamma_{\stackrel{\textstyle +}{-}}$ are the eigenvalues of the
matrix $\Gamma_{ab}$ and $\bar{z}^2=z^2e^{-h(\gamma)}$ whose
asymtotic expressions read $\Gamma_+= N ln\frac{s}{M^2}$,
$\Gamma_-=\pi^2{N^2-1\over N^3}\frac{1}{ln{s\over M^2}}$.

Given that
\begin{eqnarray}
\int_\lambda ^{\frac{1}{\bar{z}}}\frac{d\tau}{\tau}\,\frac{\alpha_s}{\pi}
&=&\frac{2}{\beta_o}\, ln\left [\frac{ln(\frac{1}{\bar{z}\Lambda^2_{QCD}}}
{ln(\frac{\lambda^2}{\Lambda^2_{QCD}}}\right ]\nonumber\\ &=&\quad
\frac{2}{\beta_o} ln\left [1\,+\,\frac{\beta_o}{4\pi}\alpha_s
ln\frac{1}{z^2\hat{\lambda}^2}\right ],
\end{eqnarray}
where $\hat{\lambda}^2=\lambda^2\frac{m^2}{s}$, we obtain
\begin{equation}
T_{\stackrel{\textstyle +}{-}}=2\alpha_s\, exp\left [
-\frac{\alpha_s}{2\pi} \Gamma_{\stackrel{\textstyle +}{-}}
ln\frac{(-t)}{\hat{\lambda}^2}\right ] \Gamma(1+\frac{\alpha_s}{2\pi}
\Gamma_{\stackrel{\textstyle +}{-}})/\Gamma(1-\frac{\alpha_s}{2\pi}
\Gamma_{\stackrel{\textstyle +}{-}}).
\end{equation}

The above result contributes to the octet part of the forward amplitude
through an expression which explicitly exhibits the reggeization of the
exchanged gluons:
\begin{equation}
T_{LL}\,\sim\, \left (\frac{s}{M^2}\right )^\beta,
\end{equation}
where $LL$ stands for `leading logarithm' and
\begin{equation}
\beta\,=\,-\frac{\alpha_s}{2\pi}N ln\frac{-t}{\hat{\lambda}^2} \, =\,
-\frac{\alpha_s}{2\pi}\left[ln\frac{-t}{\lambda^2}+ln\frac{s}{m^2}
\right],
\end{equation}
implying the more suggestive form
\begin{equation}
T_{LL}\,\sim\, \left(\frac{s}{m^2}\right)^{\alpha (t)}\,
e^{-\frac{\alpha_s}{2\pi}N ln^2\frac{s}{m^2}}.
\end{equation}
One reads, from the above expression, the Regge trajectory as $\alpha
(t)=-\frac{\alpha_s}{2\pi}N
ln\frac{-t}{m^2}$.

A notable difference is recorded with respect to gluon reggeization
results obtained by other, non-worldline, methodologies, namely the
appearance of the exponential factor $e^{-{\alpha_s\over 2\pi}Nln^2{s\over
m^2}}$. This is directly attributable to our use of an off mass
shell IR regularization strategy, as opposed to the on mass shell practice
employed in other works. Comparing, e.g., with Ref. 29, where the
anomalous dimension structure for quark scattering was first investigated,
one observes full agreement with our results. In particular, modulo a
reverse designation of the $+$ and $-$ components, the eigenvalues of the
anomalous dimension matrix coincide.

Our final expression for the amplitude, with its damping factor, presents
an inetrest of its own in connection with unitarity requirements. The
general guidelines for effecting unitarization in the amplitudes for high
energy processes, in the conventional framework of Feynman diagrams,  
have been elegantly discussed by Cheng and Wu (Ref. 1, last chapter). On a
more concrete basis, systematic attempts to deal with unitarization
of the quark-quark scattering amplitude have been pursued by Lipatov
[29-31] who has confronted the unitarity issue, within the context of
multi-Regge kinematics, in terms of an eikonal-based expression for the
S-matrix in the impact parameter space.

Even though we shall not enter unitarization issues in the present paper,
it is worth making some comparisons with more recent studies, [32-34],
which employ similar methods with ours to arrive at a description of 
high-energy scattering
in QCD in terms of an effective two-dimensional field theory. Focusing on
unitarity and gauge invariance, the above authors have recognized the
importance of facilitating the derivation of such effective actions by 
employing Wilson, straight-line contour integrals. In our approach, of
course, Wilson line operators are an integral part of the very formulation
of the field system and carry, in fact, its dynamics. The main difference
is that whereas we rely on Wilson lines of {\it finite}
extent, in the work of Refs. 32-34 lines of {\it infinite} extent,
equivalently Wilson loops, are employed. The resulting off mass shell
treatment of IR divergences in our case offers a different perspective in
that it differentiates what is `soft', but observable, and what is
`infrared' and attributable to unobservable, with respect to the
scattering dynamics, modes (wavelengths $\geq {1\over\sigma}$). The latter
contribute exclusively to the
non-perturbative dynamics of QCD and we surmise that their exclusion
from our considerations is precisely the reason for the emergence of the 
damping factor in (34). It is certainly of great interest to identify the
connection between the conventional multi-production, in the $s$-channel, 
approach to unitarity and the damping factor which makes its appearance in
our work.

\vspace*{.2in}
{\bf 2.5 Miscellaneous remarks}

A number of observations and/or remarks stemming from our worldline 
approach to non-abelian scattering in the Regge limit and which might be
of some interest will be presented in this subsection.

To begin, we wish to
consider possible connections with past work centered around Wilson loops
[20,21]. To this end, let us focus on the on-mass shell case where the
matter particle worldlines extend to infinity. In a Euclidean space-time
background two such lines join at infinity, thereby forming closed paths.
The corresponding closed loop configurations for the `crossed' and
`pair-cusp' cases are depicted, respectively, in figures 3a,b. It follows
that there is a direct correspondence between studies performed in
relation to Wilson loops and dynamical considerations taking place within
the worldline approach. The fact, on the other hand, that in our case
Wilson loops/lines enter the formulation of the field system directly and
not as formally introduced objects, underlines their role
as fundamental ingredients of the field theoretical description {\it per
se}. One immediate aftermath of this occurence has already been witnessed
in
the present work, namely the ability to utilize off-mass shell properties. 

A second point of interest concerns the relevance of the
paired-cusp configuration which entered the renormalization group study of
the forward amplitude. Cusps on Wilson loops are associated with 
bremsstrahlung radiation [19]. In a diagrammatic context, on the other
hand,
such a situation would reveal itself if we were to make a `horizontal'
unitarity-type cut  accross the $t$-channel.
It is of interest to note that the $s-t$
interplay, which is quintessential to unitarity enforcement at high
energies [1], seems to be in a one-to-one correspondence with the operator
mixing induced by the renormalization group. We feel that this is an
issue that merits further study.

Turning our attention to QED, let us observe that the abelian
nature of the theory allows us to treat the expectation value of the 
ordinary Wilson exponential as the exponential of the correlator. This
leads us directly to the eikonal form for the dynamical factor
\footnote{Aside from the obvious fact that no group indices are
involved here, there is no need for putting a subscript on $E$ as
exponentiation is now automatically obtained and a renormalization group
running is no longer required.} $E$, which, for off mass shell IR
regularization, reads
\begin{equation}
E\sim 1-e^{i\chi_o},\hspace{1cm}i\chi_o=-\frac{\alpha}{\pi} (i\pi
coth\gamma)ln{1\over z^2\hat{\lambda}^2},
\end{equation}
where $\hat{\lambda}^2\equiv\tilde{\lambda}^2e^{-h(\gamma)}$.

For the amplitude one obtains
\begin{equation}
A\sim\int d^2z e^{i\vec{q}\cdot\vec{z}}e^{i\chi_o} =4\pi i\alpha
\frac{coth\gamma}{t}\left(-{t\over\hat{\lambda}^2}\right)^{-i\alpha
coth\gamma}
\frac{\Gamma(1+i\alpha coth\gamma)}{\Gamma(1-i\alpha coth\gamma)}
\end{equation}
whose asymptotic form, as $s/m^2\rightarrow\infty$, reads
\begin{equation}
A\sim 4\pi
i\alpha\frac{1}{t}\left(-{t\over\hat{\lambda}^2}\right)^{-i\alpha}\, 
\frac{\Gamma(1+i\alpha)}{\Gamma(1-i\alpha)}\left({s\over m^2}\right)^{-i
\alpha}.
\end{equation}

One last reference to QED, which pertains to a `visual' suggestion
facilitated through its worldine casting, is the following. Suppose that 
in a
basically forward process one also allows for the observations of ``soft"
photons, i.e. photons which do not exceed a given energy scale
$\tilde{\Lambda}$. An $s$-channel study for this process can be suggested, 
in a space-time setting, by extracting a
`region' of radius $T\sim {1\over\hat{\Lambda}}$ centered around the point
of closest approach, see Fig 4a. Upon cutting, in the Feynman diagrammatic
context, along the $t$-direction we obtain
the cross sectional profile of an inclusive process involving `soft' 
photon emission, as per our requirement, see Fig 4b. The difference
brought about in (35)
corresponds to a modification of the eikonal function of the form
$\chi_o\rightarrow\tilde{\chi}_o$ which, we speculate, that for large
enough $T$ is consistent with the presence of a diffraction pattern in the
forward direction. In view of experimental observations [35] which report 
a notable excess of soft photons in the forward direction, our
aformentioned speculation might be worth to consider further.

\vspace{.4in} 
{\large {\bf 3 High-energy, non-abelian scattering at fixed angles}}

Our considerations in this Section will be extended to the case where the
four-velocities entering each of the four branches in (5) are different
from one another. In particular, we set $\dot{x}^I=u_1$ in $[0,s_1]$,
$\dot{x}^I=u'_1$ in $[s_1,T_1]$, $\dot{x}^{II}=u_2$ in $[0,s_2]$ and
$\dot{x}^{II}=u'_2$ in $[s_2,T_2]$, see Fig. 5a. For simplicity, we shall
work with
disconnected correlation functions which we denote by $(W_1)^{ii'}_{jj'}$,
where the subscript `1' pertains to the crossed configuration. (Later we 
shall use `2' as the subscript for a pair-cusped configuration which
 mixes in, under renormalization group running.)

Non-abelian group complications force us to define the following invariant
quantities, see, e.g., Ref [26],
\begin{equation}
W_1^{(a)}\equiv <trP_ItrP_{II}>_A\,=\,\delta_{ii'}\delta_{jj'}
(W_1)^{ii'}_{jj'}\nonumber\\
\end{equation}
 and
\begin{equation}
W_1^{(b)}\equiv <tr(P_IP_{II}>_A\,=\,\delta_{ij'}\delta_{ji'}
(W_1)^{ii'}_{jj'},
\end{equation}  
where $P_I$ denotes the line configuration parametrized by $\dot{x}^I$ and
$P_{II}$ the one parametrized by $\dot{x}^{II}$. 

It follows that
\begin{equation}
(W_1)^{ii'}_{jj'}\,=\,\frac{NW_1^{(a)}-W_1^{(b)}}{N(N^2-1)}\delta_{ii'} 
\delta_{jj'}+\frac{NW_1^{(b)}-W_1^{(a)}}{N(N^2-1)}\delta_{ij'}
\delta_{ji'}.
\end{equation}

A final introductory note pertains to our kinematical parametrization. We
make the following choice [36] for the particle momenta on each of the
four branches (consistent, of course, with an over all four-momentum
conservation):
\begin{eqnarray}
& & p_1=(\sqrt{Q^2+M^2},0,0,Q),\hspace{1cm}p_2=(\sqrt{Q^2+M^2},0,0,-Q)
\nonumber\\ & & p'_1=(\sqrt{Q^2+M^2},0,Qsin\theta,Qcos\theta),
\hspace{1cm}
p'_2=(\sqrt{Q^2+M^2},0,-Qsin\theta,-Qcos\theta)
\end{eqnarray}
which, in turn, parametrizes the $s$ and $t$ variables as follows
\begin{equation}
s\,=\,(p_1+p_2)^2\,=\, 4(Q^2+M^2)
\end{equation}
and
\begin{equation}
t\,=\,(p_1-p'_1)^2\,=\, -2Q^2(1-cos\theta).
\end{equation}
The limit $s,t\rightarrow\infty$ with $s/t$ fixed will be taken in the
sense $Q\rightarrow\infty$, $\theta$ fixed.

\vspace{.2in}

{\bf 3.1 Hard-Soft Factorization in the Subtheory}

Unlike the forward scattering case we now have to face a situation where a
sizeable momentum transfer is involved in the considered process which,
according to our
parametrization, is of order $Q$. The latter sets the scale beyond 
which no corresponding degree of freedom explicitly enters our analysis,
hence it is wise to ``dress" our quantities at least down to that scale.
Within the remaining `live' sector of the theory we introduce an
intermediate scale $\Lambda$ which separates soft from hard gluons and
whose arbitrariness will naturally induce a renormalization group running 
in the subtheory. Moreover, we shall place the matter
particles on-shell, i.e. we shall employ worldlines of infinite extent,
thereby regulating the IR divergencies through a small gluon mass
$\lambda$.

As $\Lambda$ stands between $Q$ and  $\lambda$, what one calls ``soft''
and what ``hard" is relative. For example, if one were to play with
$\Lambda$, say lower it, then gluons that were originally debited to the
soft transfer to the hard group. The opposite happens, of course, when
the value of $\Lambda$ is raised. The factorized relation for the
invariant quantities $W_1^{(a,b)}$ reads
\begin{equation}
W_1^{(a,b)}\,=\,(W_1^{(a,b)})_{SOFT}(W_1^{(a,b)})_{HARD}\,+\,{\cal
O}\left({1\over\Lambda^2}\right).
\end{equation}
The arbitrariness of the dividing scale calls for a renormalization
group running which will lead to our final expression for $W_1^{(a,b)}$
and, by
extension, for the amplitude. The manner in which this strategy will be
effected is the subject of concern in the present subsection.

Let us start by recalling our discussion in subsection 2.2 according to
which, given the (cusped) line configurations for each of the two
colliding particles, soft gluons correspond to what is emitted or absorbed
by the straight line
segments (no impulse approximation). In this soft sector of the full
theory one determines anomalous dimensions associated with its own high
energy
domain\footnote{This is {\it not} a novel idea. For example, in the
Bloch-Nordsieck approximation [37], which describes the soft limit of QED,
one discovers anomalous dimensions [38] which lead to the proper form
of the full fermion propagator in the IR. The point is that, from the
perspective of the IR cutoff $\lambda$, the upper momentum scale $Q$
appears as infinite.}. For a given $\Lambda$, one can induce a
renormalization group running of $(W_1)_{SOFT}$ from $\lambda$ to
$\Lambda$.

With the above observation in place, our next remark is that
$(W_1^{(a,b)})_{SOFT}$ exhibits a dependence on $Q$ through the angle
$\theta$
formed at a given cusp, e.g. $cos\theta=u_1\cdot u'_1$. We thereby write
\begin{equation}
\frac{d}{dlnQ^2}lnW_1^{(a,b)}\,=\,\frac{d}{dlnQ^2}ln(W_1^{(a,b)})_{SOFT}+
\frac{d}{dlnQ^2}ln(W_1^{(a,b)})_{HARD}.
\end{equation}

Now, the renormalization group equation for the quantity $W_1^{(a,b)}$ as
a whole,
which runs in the interval $[\lambda,Q]$, reads
\begin {equation}
\left(\mu\frac{\partial}{\partial\mu} + \beta (g)\frac{\partial}{\partial
g}\right)\frac{d}{lnQ^2}lnW_1^{(a,b)}\,=\,0
\end{equation}
and expresses independence from the scale that separates soft from hard
physics within the considered subtheory. (We have used $\mu$ to represent
$\Lambda$ in order to underline the fact
that we are letting the latter scale to run.)

Factorization, then, gives
\begin{eqnarray}
& &\left(\mu\frac{\partial}{\partial\mu}+\beta\frac{\partial}{\partial g}
\right)\frac{d}{dlnQ^2}ln(W_1^{(a,b)})_{HARD}\nonumber\\ & &\quad =
-\left(
\mu\frac{\partial}{\partial\mu}+\beta\frac{\partial}{\partial g} \right)
\frac{d}{dlnQ^2}ln(W_1^{(a,b)})_{SOFT}.
\end{eqnarray}
But, provided we find the anomalous dimensions associated with the soft
factor, the expression on the right enters the
renormalization group equation discussed above along with a term of the
form: (anomalous dimensions)$\times (W_1^{(a,b)})_{SOFT}$. Therefore,
$W_1^{(a,b)}$ can be determined via a two-step procedure which first
addresses itself to its soft
and second to its hard component.

In the next subsection we shall carry out perturbative calculations 
pertaining to the soft part which will lead to the determination, to order
$\alpha_s$, of the anomalous dimension matrix.

\vspace{.2in}
{\bf 3.2 Perturbative calculations in the soft sector}

We begin our considerations surrounding the soft part of the amplitude  
by displaying its perturbative expression, to ${\cal O}(g^2)$, which reads
\begin{eqnarray}
\left[(W_1)_{SOFT}\right]^{ii'}_{jj'}&
&=\delta_{ii'}\delta_{jj'}+(ig)^2c_F\delta_{ii'}\delta_{jj'}\{
2\int^\infty_0d\tau\int^\infty_0 d\tau 'D(|\tau u_1-\tau
'u_1|)\nonumber\\&+&\int^\infty_0 d\tau\int^\infty_0d\tau ' u_1\cdot u_1'
D(|\tau u_1+\tau ' u_1'|)+\int^\infty_0 d\tau\int^\infty_0 d\tau '
u_2\cdot u_2' D(|\tau u_2+\tau ' u_2'|)\}\nonumber\\ &+& (ig)^2
t^a_{ii'}t^a_{jj'}\{\int^\infty_0 d\tau\int^\infty_0 d\tau ' D(|\tau
u_i-\tau ' u_2|)+\int^\infty_0d\tau\int^\infty_0d\tau ' D(|\tau u_1' +\tau
' u_2|)\nonumber\\ &+&\int^\infty_0d\tau\int^\infty_0d\tau 'u_1\cdot
u_2'D(|\tau u_1+\tau' u_2'|)+\int^\infty_0d\tau\int^\infty_0 d\tau ' u_1'
\cdot u_2'D(|\tau u_1'-\tau 'u_2'|)\}\nonumber\\&+&{\cal O}(g^4)
\end{eqnarray}
with $D(|x|)$ given by Eq. (22) (we have suppressed the $\mu$ argument for
simplicity).  

We determine
\begin{equation}
\int^\infty_0d\tau\int^\infty_0d\tau 'D(|\tau u_1+\tau 'u_2|)=
\frac{1}{(4\pi)^{D/2}}\left(\frac{\mu}{\lambda}\right)^{4-D} 2\Gamma(2-
\frac{D}{2})\frac{1}{\sqrt{1-w^2}}arctg\frac{\sqrt{1-w^2}}{w},
\end{equation}
where $w\equiv u_1\cdot u_2$ and
\begin{equation}
\int^\infty_0d\tau\int^\infty_0d\tau 'D(|\tau u_1-\tau 'u_2|)=
\frac{1}{(4\pi)^{D/2}}\left(\frac{\mu}{\lambda}\right)^{4-D} 2\Gamma(2-
\frac{D}{2})\frac{1}{\sqrt{1-w^2}}\left[\pi
-arctg\frac{\sqrt{1-w^2}}{w}\right].
\end{equation}

The above relations together with (9) give, upon transcription to
Minkowski space,
\begin{eqnarray}
[(W_1)_{SOFT}]^{ii'}_{jj'}&=&\delta_{ii'}\delta_{jj'}-{g^2\over 4\pi^2}
\left(\frac{\mu^2}{\lambda^2}\pi\right)^\epsilon\frac{\Gamma(1+\epsilon)}
{\epsilon}\left\{c_F\delta_{ii'}\delta_{jj'}(\gamma_{11'} coth\gamma_{11'}
-1)\right. \nonumber\\&+&\left. 
t^a_{ii'}t^a_{jj'}[(i\pi-\gamma_{12}coth\gamma_{12}
+\gamma_{12'}coth\gamma_{12'}]\right\}+{\cal O}(g^4),
\end{eqnarray}
where $coth\gamma_{ij}={1\over m^2}p_i\cdot p_j$ and where we have taken
into account that $\gamma_{12}=\gamma_{1'2'},\,\gamma_{12'}=\gamma_{1'2}$,
due to momentum conservation.

For the corresponding invariant quantities $W_1^{(a,b)}$ we find
\begin{equation}
(W_1^{(a)})_{SOFT}=\,N^2\left[1-\frac{\alpha_s}{\pi}ln\left(\frac{\mu}{\lambda} 
\right)A_{11}\right]-N\frac{\alpha_s}{\pi}ln\left(\frac{\mu}{\lambda}\right)A_{12}
+{\cal O}(\alpha_s^2)
\end{equation}
and
\begin{equation}
(W_1^{(b)})_{SOFT}=\,N\left[1-\frac{\alpha_s}{\pi}ln\left(\frac{\mu}{\lambda}
\right)A_{11}\right]-N^2\frac{\alpha_s}{\pi}ln\left(\frac{\mu}{\lambda} 
\right)A_{12}+{\cal O}(\alpha_s^2),
\end{equation}
where
\begin{equation}
A_{11}=2c_F(\gamma_{11'}coth\gamma_{11'}-1) - {1\over N}\left[(i\pi
-\gamma_{12})coth\gamma_{12}+\gamma_{12'}coth\gamma_{12'}\right]
\end{equation}
and
\begin{equation}
A_{12}=(i\pi
-\gamma_{12})coth\gamma_{12}+\gamma_{12'}coth\gamma_{12'}.
\end{equation}

We now bring into play the quantity $[(W_2)_{SOFT}]^{ii'}_{jj'}$
given by
\begin{eqnarray}
[(W_2)_{SOFT}]^{ii'}_{jj'}&=&<Pexp[ig\int_{-\infty}^0d\tau u_1\cdot
A(\tau u_1)+ig\int^{\infty}_0d\tau u_2'\cdot A(\tau
u_2')]_{ij'}\nonumber\\& &\quad\times Pexp[ig\int_{-\infty}^0d\tau
u_2\cdot
A(\tau u_2)+ig\int^{\infty}_0d\tau u_1'\cdot A(\tau
u_1')]_{ji'}>_A
\end{eqnarray}
which mixes with $[(W_1)_{SOFT}]^{ii'}_{jj'}$ under the renormalization
group. The relevant configuration is depicted in Fig. 5b.

Similar considerations to those that led to (51) now give (in Minkowski
space) 
\begin{eqnarray}
[(W_2)_{SOFT}]^{ii'}_{jj'}& &=\delta_{ij'}\delta_{ji'}-{g^2\over 4\pi^2}
\left(\frac{\mu^2}{\lambda^2}\pi\right)^\epsilon\frac{\Gamma(1+\epsilon)}
{\epsilon}\left\{c_F\delta_{ij'}\delta_{ji'}(\gamma_{12'} coth\gamma_{12'}
-1)\right.\nonumber\\&+&\left.
t^a_{ij'}t^a_{ji'}[(i\pi-\gamma_{12}coth\gamma_{12}
+\gamma_{11'}coth\gamma_{11'}]\right\}+{\cal O}(g^4).
\end{eqnarray}    
The corresponding invariant quantities $(W_2^{(a,b)})_{SOFT}$ turn out to
be
\begin{equation}
(W_2^{(a)})_{SOFT}=\,-N^2\frac{\alpha_s}{\pi}ln\left(\frac{\mu}{\lambda}
\right)A_{21}+N\left[1-\frac{\alpha_s}{\pi}ln\left(\frac{\mu}{\lambda}\right) 
A_{22}\right]+{\cal O}(\alpha_s^2)
\end{equation}
and
\begin{equation}
(W_2^{(b)})_{SOFT}=\,-N\frac{\alpha_s}{\pi}ln\left(\frac{\mu}{\lambda}
\right)A_{21}+N^2\left[1-\frac{\alpha_s}{\pi}ln\left(\frac{\mu}{\lambda} 
\right)A_{22}\right]+{\cal O}(\alpha_s^2),
\end{equation}
where
\begin{equation}
A_{21}=(i\pi
-\gamma_{12})coth\gamma_{12}+\gamma_{11'}coth\gamma_{11'}
\end{equation}
and
\begin{equation}
A_{22}=2c_F(\gamma_{12'}coth\gamma_{12'}-1) - {1\over N}\left[(i\pi
-\gamma_{12})coth\gamma_{12}+\gamma_{11'}coth\gamma_{11'}\right].
\end{equation}
With the above results in place, we are ready to apply the 
renormalization group analysis for the fixed angle scattering amplitute.
The relevant presentation will be given in the next subsection.

\vspace{.2in}
{\bf 3.3 Renormalization Group running and Sudakov behaviour}

Our perturbative results, to ${\cal O}(\alpha_s)$, of the previous
subsection lead to a LLA for $(W_1^{(a,b)})_{SOFT}$ via the
renormalization group (RG) equation
\begin {equation}
\left(\mu\frac{\partial}{\partial\mu} + \beta (g)\frac{\partial}{\partial
g}\right)\tilde{W}_{SOFT}^{(i)}\,=\,-\frac{\alpha_s}{\pi}\tilde{A}\, 
\tilde{W}_{SOFT}^{(i)},\hspace{.1in}i=a,b,
\end{equation} 
where
\begin{equation}
\tilde{W}_{SOFT}^{(i)}\equiv \left(\begin{array}{c}W_1^{(i)}\\ W_2^{(i)}
\end{array}\right)_{SOFT}
\end{equation}
and
\begin{equation}
\tilde{A}\,=\,\left(\begin{array}{cc}A_{11}\,\,&\,\,A_{12}\\A_{21}\,\,&\,\,
A_{22}
\end{array}\right).
\end{equation}

The boundary conditions for solving the RG equation are chosen so that no
structure is seen at momentum scales below the IR cutoff $\lambda$:
\begin{equation}
\left.\tilde{W}_{SOFT}^{(a)}\right|_{\mu=\lambda}=\left(\begin{array}{c}N^2\\N
\end{array}\right),\hspace{.1in}\left.\tilde{W}_{SOFT}^{(b)} 
\right|_{\mu=\lambda}=\left(\begin{array}{c}N\\N^2
\end{array}\right). 
\end{equation}

The solution has the general form
\begin{equation}
\tilde{W}_{SOFT}^{(i)}(\mu/\lambda)=Pexp\left[-\tilde{A}\int_\lambda^\mu
\frac{d\tau}{\tau}\frac{\alpha_s(\tau)}{\pi}\right]\tilde{W}_{SOFT}^{(i)}(1).
\end{equation}

In the asymptotic regime of interest we determine
\begin{equation}
\gamma_{12}=cosh^{-1}\left(\frac{s}{2m^2}-1\right)=
cosh^{-1}\left(\frac{2Q^2}{m^2}+1\right) \simeq ln\left(\frac{2Q^2}{m^2}
\right)
\end{equation}
and, in a similar fashion,
\begin{equation}
\gamma_{11'} \simeq
ln\left(\frac{2Q^2}{m^2}\right)+ln(sin^2\theta),\hspace{1cm}
\gamma_{12'} \simeq ln\left(\frac{2Q^2}{m^2}\right)+ln(cos^2\theta).
\end{equation}

Taking the above into account and following the procedure exhibited in the
Appendix we arrive at the asymptotic, as $s,|t|\rightarrow\infty$ at fixed
ratio, results
\begin{equation}
(W_1^{(a)})_{SOFT}\simeq N^2exp\left[-2c_Fln\left({2Q^2\over m^2}\right)
\int_\lambda^\mu \frac
{d\tau}{\tau}{\alpha_s(\tau)\over\pi}\right]4c_Fln\left({2Q^2\over
m^2}\right)\phi_1(\mu/\lambda,\theta)
\end{equation}
and
\begin{equation}
(W_1^{(b)})_{SOFT}\simeq Nexp\left[-2c_Fln\left({2Q^2\over m^2}\right)
\int_\lambda^\mu \frac
{d\tau}{\tau}{\alpha_s(\tau)\over\pi}\right]4c_Fln\left({2Q^2\over
m^2}\right)\phi_1(\mu/\lambda,\theta),
\end{equation}
where the function $\phi_1(\mu/\lambda,\theta)$ is defined in the
Appendix, along with a function $\phi_2(\mu/\lambda,\theta)$ which does
not appear in the above expressions since its role is inconsequential to
our subsequent considerations. Note that ${1\over
N}(W_1^{(a)})_{SOFT}\simeq (W_1^{(b)})_{SOFT}$.

The resulting expression for the amplitude is
\begin{equation}
[(W_1)_{SOFT}]^{ii'}_{jj'}\simeq\,[\delta_{ii'}\delta_{jj'}+\frac{N}{N^2-1} 
\delta_{ij'} \delta_{ji'}]F_{SOFT}\left({Q^2\over m^2},{\mu\over\lambda}
\right),
\end{equation}
where
\begin{equation}
F_{SOFT}\left({Q^2\over m^2},{\mu\over\lambda}\right)=4c_F
ln\left({2Q^2\over m^2}\right)exp\left[-2c_Fln\left({2Q^2\over m^2}\right)
\int_\lambda^\mu \frac
{d\tau}{\tau}{\alpha_s(\tau)\over\pi}\right]\phi_1(\mu/\lambda,\theta).
\end{equation}

According, now, to the guidelines set by our discussion in subsection 3.1
we proceed to determine that
\begin{equation}
\frac{d}{dlnQ^2}ln(W_1^{(i)})_{SOFT}=-2c_Fln\left({2Q^2\over m^2}\right)
\int_\lambda^\mu \frac{d\tau}{\tau}\frac{\alpha_s(\tau)}{\pi} +{1\over
lnQ^2}
\end{equation}
whereupon, with the aid of (47), we deduce
\begin{equation}
\frac{d}{dln\mu}\frac{d}{dlnQ^2}ln(W_1^{(i)})_{HARD}={2c_F\over\pi}
\alpha_s(\mu),
\end{equation}
or
\begin{equation}
\frac{d}{dlnQ^2}ln(W_1^{(i)})_{HARD}=-2c_F\int_\mu^{|Q|}
\frac{d\tau}{\tau}\frac{\alpha_s(\tau)}{\pi}+{\cal R}(\alpha_s(Q)).
\end{equation}
The above result when put together with (73) gives
\begin{equation}
\frac{d}{dlnQ^2}ln(W_1^{(i)})=-2c_F\int^{\mu^2}_{\lambda^2} 
\frac{d\tau}{2\tau}\frac{\alpha_s(\tau)}{\pi}-2c_F\int_{\mu^2}^{Q^2}
\frac{d\tau}{2\tau}\frac{\alpha_s(\tau)}{\pi}+{\cal R}(\alpha_s(Q)),
\end{equation}  
which leads to
\begin{equation}
W_1^{(i)}\,=\,{\cal L}[\alpha_s(Q^2)]\,exp\left[-2c_F\int_{\mu^2}^{Q^2}
\frac{d\tau}{2\tau}\frac{\alpha_s(\tau)}{\pi}ln{Q^2\over\tau}\right].
\end{equation}
But
\begin{eqnarray}
2c_F\int_{\mu^2}^{Q^2}
\frac{d\tau}{2\tau}\frac{\alpha_s(\tau)}{\pi}ln{Q^2\over\tau}&=&
\frac{4c_F}{\beta_o}\left[ln\left({Q^2\over\Lambda^2}\right)
-lnln\left({Q^2\over\Lambda^2}\right)\right.\nonumber\\ & &\quad \left.
ln\left({Q^2\over\Lambda^2}\right)lnln\left({\lambda^2\over\Lambda^2}\right)
-ln\left({Q^2\over\Lambda^2}\right)\right],
\end{eqnarray}
which explicitly exhibits Sudakov behaviour for the $W_1^{(i)}$
through its leading term [36].

Some comments are in order at this point. Beginning with technical issues,
let us first notice that, in the asymptotic regime under consideration, 
$W_1^{(a)}\sim W_1^{(b)}$. This means that the Sudakov behaviour of the
$W_1^{(i)}$ passes on to the full expression for
$(W_1)^{ii'}_{jj'}$. Moreover, nothing changes if one goes to the
corresponding quantity $(E_1)^{ii'}_{jj'}$ associated with the connected
four-point function since the two disjoint cusped line configurations,
which make
the difference between the connected and disconnected expressions, will
each provide similarly suppressing Sudakov form factors. Finally, particle
exchange in the final states, which cannot be {\it a priori} excluded from
consideration in the fixed angle case, simply permutes the $s$ with the
$t$ variable without affecting our results. Let us also note, on
the technical front, that our final expressions contain non-leading terms
whose assesment should provide interesting new information.

From a physical standpoint the message to be drawn from the Sudakov
behaviour that has been extracted for the fixed angle amplitude amounts to
the standard realization that the larger the momentum transfer between the
colliding particles the smaller the probability for the process to remain
exclusive.

\vspace{.4in}
{\large {\bf 4 Concluding Remarks}}

The worldline casting of gauge theories with spin-1/2 matter fields has as
its basic feature the space-time setting within which physical
quantities are described. Both particle propagation and dynamics, the
latter in the form of Wilson lines, are embodied in space-time paths.
Generically, of course, all possible contours enter the path integral. By
restricting ourselves to paths that are straight almost everywhere, we
were able to achieve a sharp factorization of a soft, relevant to the
process, sector {\it at the fundamental field theoretical level}.

For the case where no cusps, to break the straight contours, are present
we are dealing with situations where the no impulse approximation holds
throughout, equivalently the `soft' subsector represents 
the full field theory. Nevertheless, {\it it is} possible to determine
anomalous dimensions governing processes in this domain, an
occurence which reflects the fact that from the viewpoint of the IR cutoff
$\lambda$ the `upper roof' $\Lambda$ of the soft subsector appears to be
infinite. For the
four-point, forward scattering process considered in this paper the
corresponding RG considerations led to gluon reggeization.

For processes in which cusped configurations make their entrance a
non-negligible momentum transfer takes place on the basis of a sudden
impulse approximation. This time the RG running aquires the standard
interpretation of a factorization between soft and hard physics, within
the isolated, with respect to the considered energy range, subtheory. This
situation is analogous to the operator product
expansion that separates Wilson coefficients (hard factors) from operator
expectation values (soft factors). In the fixed angle scattering regime
that we considered in section 3, the end result was the emergence of
Sudakov behavior for the amplitude.

We hope to have sufficiently illustrated the efficiency by which the
factorization of soft physics can be attained within the worldline
casting of non-abelian gauge theories. Morever, once familiarization with
computational methodology and procedure is aquired, one realizes that the
worldline handling of soft subsectors involves more or less similar
mathematical expressions, irrespective of the process one studies. Thus,
along
with the conceptual simplicity regarding the factorization strategy there
are additional advantages, of practical nature, to the proposed approach
as well. 

Clearly, the results we have exhibited are valid to ${\cal O}(\Lambda)$.
In the OPE language this amounts to leading twist. Non leading
contributions are lurking in our expressions and we intend to study their
implications in future work. More interesting is the question
concerning the
relation between the factorization advocated in this paper and the
standard factorization widely discussed in the literature with basic 
reliance on Feynman diagrammatic logic,  especially in
connection with exclusive processes. We intend to report on this issue in
a forthcoming paper. 

\newpage

\appendix
\setcounter{section}{0}
\addtocounter{section}{1}
\section*{Appendix}

\setcounter{equation}{0}
\renewcommand{\theequation}{\thesection.\arabic{equation}}

We trace the steps which take us from (66) to (69) and (70).

We define
\begin{equation}
\hat{C}(\tilde{A})\equiv
exp\left[-\tilde{A}\int_\lambda^\mu\frac{d\tau}{\tau}
\frac{\alpha_s(\tau)}{\pi}\right],\nonumber\\
\end{equation}
where $\tilde{A}$ stands for the $2\times 2$ matrix given by (64).

The following identity holds
\begin{eqnarray}
\hat{C}(\tilde{A})&=&\frac{A_+-\tilde{A}}{A_+-A_-}\hat{C}(A_-)- 
\frac{A_--\tilde{A}}{A_+-A_-}
\hat{C}(A_+)\nonumber\\&=&\frac{1}{A_+-A_-}
[A_+\hat{C}(A_-)-A_-\hat{C}(A_+)]+\tilde{A}\frac{1}{A_+-A_-}
[\hat{C}(A_+)-\hat{C}(A_-)],\nonumber\\
\end{eqnarray}
where $A_{\stackrel{+}{-}}$ are the eigenvalues of $\tilde{A}$.

We write
\begin{equation}
\hat{C}(\tilde{A})_{11}=X+YA_{11},\hspace{1cm}\hat{C}(\tilde{A})_{12}
=YA_{12},\nonumber\\
\end{equation}
where
\begin{equation}
X\equiv \frac{1}{A_+-A_-}
[A_+\hat{C}(A_-)-A_-\hat{C}(A_+)],\hspace{.5cm}Y\equiv \frac{1}{A_+-A_-}
[\hat{C}(A_+)-\hat{C}(A_-)].\nonumber\\
\end{equation}

It follows from Eq. (66) in the text, that
\begin{equation}
(W_1^{(a)})_{SOFT}=N^2X\,+\,YN(NA_{11}+A_{12})\nonumber\\
\end{equation}
and
\begin{equation}
(W_1^{(b)})_{SOFT}=NX\,+\,YN(A_{11}+NA_{12}).\nonumber\\
\end{equation}

Referring to Eqs. (54), (55), (60), (61) in the text and taking into
account the asymptotic conditions (67, (68) we determine
\begin{eqnarray}
& &A_{11}\simeq
2c_F\,ln\left(\frac{2Q^2}{m^2}\right)+2c_F\,ln(sin^2\frac{\theta}{2})
-\frac{1}{N}\,ln(cos^2\frac{\theta}{2})-\frac{i\pi}{N}\nonumber\\
& &A_{12}\simeq i\pi +ln(cos^2\frac{\theta}{2})\nonumber\\ & &
A_{21}\simeq i\pi +ln(sin^2\frac{\theta}{2})\nonumber\\ & & A_{22}\simeq  
2c_F\,ln\left(\frac{2Q^2}{m^2}\right)+2c_F\,ln(cos^2\frac{\theta}{2})
-\frac{1}{N}\,ln(sin^2\frac{\theta}{2})-\frac{i\pi}{N}.
\end{eqnarray}
 We therby obtain
\begin{eqnarray}
A_{\stackrel{+}{-}}&\simeq &2c_F\,ln\left(\frac{2Q^2}{m^2}\right)+
(c_F-\frac{1}{2N})ln({1\over 4}sin\theta)-{i\pi\over
N}\nonumber\\&\stackrel{+}{-}&{1\over 2}\sqrt{N^2ln^2({1\over
4}sin\theta)-4\pi^2+4i\pi sin({1\over 4}sin\theta)}.
\end{eqnarray}
Substituting into the expressions for $X$ and $Y$ we find
\begin{equation}
X=exp\left[-2c_Fln\left({2Q^2\over m^2}\right)
\int_\lambda^\mu \frac
{d\tau}{\tau}{\alpha_s(\tau)\over\pi}\right]\left\{2c_Fln\left({2Q^2\over
m^2}\right)\phi_1(\mu/\lambda,\theta)+\phi_2(\mu/\lambda,\theta)\right\}
\end{equation}
and
\begin{equation}
Y=exp\left[-2c_Fln\left({2Q^2\over m^2}\right)
\int_\lambda^\mu \frac
{d\tau}{\tau}{\alpha_s(\tau)\over\pi}\right]\phi_1(\mu/\lambda,\theta),
\end{equation}
where
\begin{equation}
\phi_1(\mu/\lambda,\theta)\equiv\frac{1}{C_+-C_-}\left[e^{-C_-\int_\lambda^\mu
\frac{d\tau}{\tau}\frac{\alpha_s(\tau)}{\pi}}-e^{-C_+\int_\lambda^\mu
\frac{d\tau}{\tau}\frac{\alpha_s(\tau)}{\pi}}\right]
\end{equation}
and
\begin{equation}
\phi_2(\mu/\lambda,\theta)\equiv\frac{1}{C_+-C_-} 
\left[C_+e^{-C_-\int_\lambda^\mu
\frac{d\tau}{\tau}\frac{\alpha_s(\tau)}{\pi}}-C_-e^{-C_+\int_\lambda^\mu
\frac{d\tau}{\tau}\frac{\alpha_s(\tau)}{\pi}}\right]
\end{equation}

Substituting into the relations giving the $(W_1^{(i)})_{SOFT}$ and
keeping only  the $Q^2$-dependent part of the resulting expressions we
finally arrive at Eqs. (69) and (70) given in the text.

\newpage

\begin{center}
FIGURE CAPTIONS
\end {center}

{\bf Fig. 1a}: A Cheng-Wu tower entering the (forward) fermion-fermion
scattering process in QED. The unitarity cut (dashed line) reveals a
fragmented profile of an electron (positron).

{\bf Fig. 1b}: Depiction of a forward, high-energy fermion-fermion
scattering process in a non-abelian gauge field theory. Gluons linking the
scattered particles reggeize while the fragmented profile of the latter is
presented by gluons exchanged between reggeons.

\vspace{.5cm}

{\bf Fig. 2a}: Worldline depiction of forward elastic scattering between
spin-1/2 matter particles. Closest distance of approach (impact parameter)
is $z$.

{\bf Fig. 2b}: Double cusped configuration which mixes with the four-point
function associated with the forward scattering process, depicted in Fig.
2a, under renormalization group running.

\vspace{.5cm}

{\bf Fig. 3a}: Wilson loop version of Fig. 2a, corresponding to an on mass
shell situation.

{\bf Fig. 3b}: Wilson loop version of Fig. 2b, corresponding to an on mass
shell situation.

\vspace{.5cm}

{\bf Fig. 4a}: Worldline depiction of a (near) forward process which
excludes photon exchange within a region of size $T$ around the point of
closest approach.

{\bf Fig. 4b}: Feynman diagrammatic representation of the situation
depicted in Fig. 4a.

\vspace{.5cm}

{\bf Fig. 5a}: Worldline depiction of fermion-fermion scattering at fixed
angle, in the sudden implulse approximation.

{\bf Fig. 5b}: Worldline depiction of the, double-cusped, contour
associated with the operator that mixes with the fixed angle scattering
one, under renormalization group running.

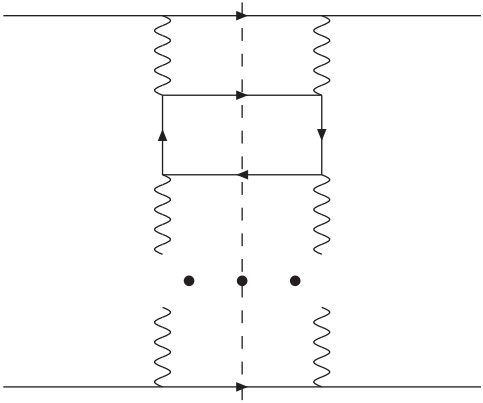
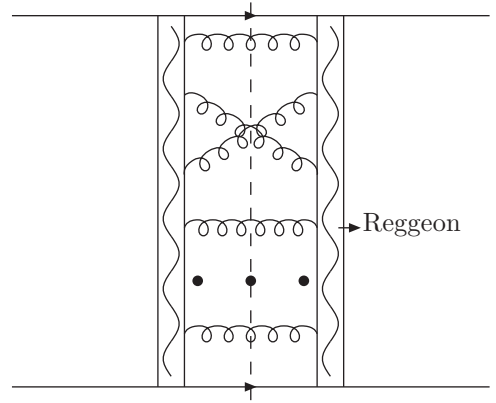
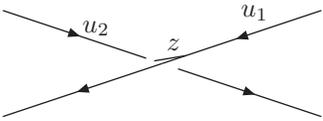
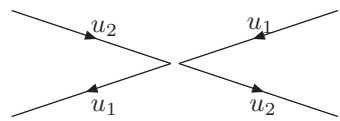
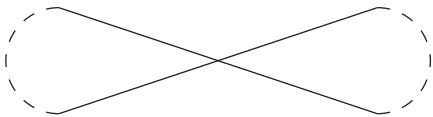
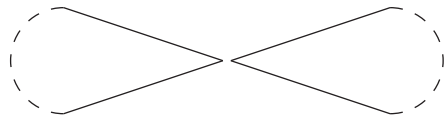
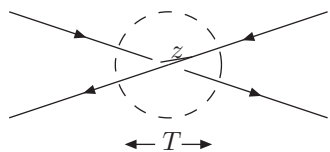
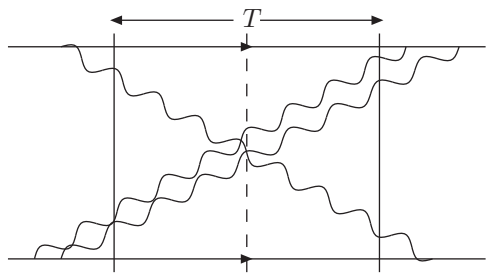
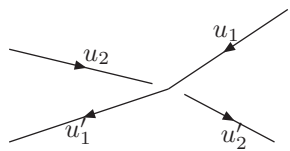
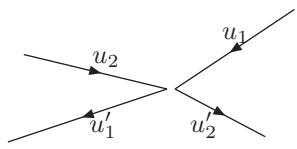
\end{document}